\documentclass[showpacs,aps,prl]{revtex4}
 
\usepackage{graphicx,amsmath}
\usepackage[dvips]{color}
\usepackage[normalem]{ulem} 
\newcommand{\beq}{\begin{equation}}
\newcommand{\eeq}{\end{equation}}
\newcommand{\bea}{\begin{eqnarray*}}
\newcommand{\eea}{\end{eqnarray*}}
\newcommand{\bnea}{\begin{eqnarray}}
\newcommand{\enea}{\end{eqnarray}}
\newcommand{\p}{\partial}
\newcommand{\ba}{\begin{array}{cc}}
\newcommand{\ea}{\end{array}}
\newcommand{\bm}{\left[ \begin{array}{cc}}
\newcommand{\fm}{\end{array} \right]}

\begin{document}

\title{Signal amplification and control 
in optical cavities with off-axis feedback}

\author{Roberta Zambrini}
\affiliation{Cross-Disciplinary Physics Department IMEDEA (CSIC-UIB), Palma
de Mallorca, 07122, Spain.}
\email{roberta@imedea.uib.es}
\author{Francesco Papoff}
\affiliation{Department of Physics, University
of Strathclyde,  107 Rottenrow,  Glasgow G4 0NG, UK.}
\date{\today}

\begin{abstract}

We consider a large class of optical cavities and gain media with  an off-axis
external feedback which introduces a two-point nonlocality.  This nonlocality
moves the lasing threshold and opens large windows of control
parameters where  weak light spots can be strongly amplified while the
background  radiation remains very low. Furthermore, transverse phase and group
velocities of a signal can be independently tuned and this enables to steer it
non mechanically, to control its spatial chirping and to split it into
two counter-propagating ones.

\end{abstract}

	
\pacs{42.65.Sf, 42.55.-f, 89.75.Kd} 

\maketitle

In this letter we study the effects introduced by a two-point nonlocality 
\cite{arecchi} on a broad class of nonlinear equations with both diffusion and
diffraction. Systems modelled by this type of equations can be experimentally
realised in optics  by cavities with an off-axis external  feedback, which is
the spatial  analogous of a feedback with temporal delay \cite{pere}. Off-axis
feedback has been subject of theoretical and experimental study in liquid
crystals light valves \cite{ramazza}, Kerr-like media \cite{thorsten,taki} and
generic nonlinear systems with diffusive coupling \cite{papoff05a}. Here instead
we consider a broad class of {\sl optical cavities} using a general formalism that can
be applied to gases, solid state and semiconductor media with fast decay of the
polarization, including media with negative refractive index and devices with
soft apertures.  The simultaneous presence of diffusive and diffractive terms 
appears in universal Ginzburg-Landau equations describing the behaviour of  any
spatially extended system near the onset of oscillations, such as,  for
instance, reaction-diffusion systems and lasers \cite{cross93}.   These
equations describe also properties of systems with time-delayed feedback and no
spatial degrees  of freedom when the delay is order of magnitude larger than the
other  time scales \cite{giacomelli98}, with the  slow time formally taking the
role of the spatial variable.\\
We show that the inclusion of a two-point nonlocality generalises these equations 
introducing new regimes and is a powerful way to amplify, characterise and control
perturbations, either external or intrinsic to the system. In particular,
nonlocality changes the nature of the first instability, which without
nonlocality leads to a spatially extended, lasing state.  With nonlocality, on
the contrary, there are large windows of control parameters where  small
localized signals can be strongly amplified while the background  radiation in
other region of the system remains very low. Furthermore, the signal moves 
across the  cavity with transverse phase and group velocities that are easily 
managed to have the same or opposite signs.    It is indeed possible, {\sl
without altering the mechanical alignment} of the set-up, to control signals
motion, tuning continuously the group velocity so that a localized perturbation
is steered either towards  or against the off-set direction and can even be
{\sl split} into two counter-propagating components. The tunability of the phase
velocities allows to control the spatial chirping of light signals independently
from the direction of steering. These unusual properties open new possibilities
for light control and can underpin
applications in optical communications, imaging and micromanipulation.

In the following we analyse how the first threshold depends on nonlocality,
diffusion and diffraction, determine the nature of the
instability, find a second threshold and  derive the
equations for the phase and group velocity of localized  perturbations. We
consider optical systems described by non-dimensional equations  of the type 
\bnea 
\p_t E &=& g_1(|E|^2,N; \mu)E + e^{i\delta} \p^2_{xx} E + r e^{i
\phi}  E_{\Delta x} ,\label{eqs_classB} \\ \nonumber \p_t N &=&
g_2(|E|^2,N,\p^2_{xx}N;\mu),   
\enea 
where $E$ is the slowly-varying amplitude of the electric field, $N$ is the
population inversion and $\mu$ is a control parameter.  We consider here one
transverse dimension $x$ as nonlocality changes only the spatial dependence of the
dispersion along the direction of the shift. Time and space are scaled with field
decay and with the square root of the modulus of the Laplacian coefficient. Our
analysis encompasses devices with diffusion that is due to Fourier filtering by
intracavity soft apertures \cite{dunlop97b}  or to elimination of the fast 
variables \cite{coullet89a}, as well as media
with positive or negative refractive index \cite{neg_refr}.  $\delta$ gives the
relative strength of diffusion and diffraction, with $\delta \in (0,\pi/2)$   for
positive  refractive indexes and $\delta \in (-\pi/2,0)$ for negative indexes,
corresponding to  left-handed materials. The term  $r e^{i \phi} E_{\Delta x}$
represents  nonlocal coupling of the field $E$  in a point $x$ with the field 
$E_{\Delta x}$  in a point $x+ \Delta x$ and is the consequence of an off-axis,
single-passage  feedback loop.  This is characterised by an amplitude $0<r<1$ and
a phase shift $\phi$ accumulated by the fast component of the electric  field in
the external loop. We assume here that the temporal  delay of the feedback is
negligible compared to the time scales of $E$ and $N$. The generic complex
functions $g_{1,2}$  allow us to describe all class B lasers, including
semiconductor. The following analysis  immediately applies also to (i) the simpler
case of systems in which the variable $N$ can be eliminated (class A) and (ii)  a
more general class of equations in which the feedback term is nonlinear
\cite{footnote2}.

We consider perturbations $\delta E \propto \exp{(\omega t +i k x)}$
of the non lasing solution $E_0=0$ and $N_0$ such that $g_2(0,N_0) =0$.
These perturbations have complex dispersion  relation  
\beq \omega = \beta - e^{i
\delta} k^2 +r e^{i (\phi +k \Delta x)}, \label{ap_disp}
\eeq 
with $\beta=g_1(0,N_0;\mu)$ also complex. 
In the following real and imaginary parts of complex quantities have subindices
$R$ and $I$, respectively.
In the limit of vanishing shift $\Delta x=0$, 
the laser threshold, given by $\beta_R^{th}= -r \cos (\phi )$, decreases when
the feedback interferes constructively with the intracavity field and increases
when the interference is destructive. Because the fast relaxation of the
polarization implies that the gain bandwidth is very large, 
all travelling waves
have the same gain/loss if there is no diffusion. 
The effect of diffusion is to filter the high Fourier components so that the
most unstable mode
is the homogeneous
one ($k=0$) independently from the relative strength of diffusion and
diffraction ($\delta$).
When $\Delta x \ne 0$, on the other hand, the most unstable mode can have $k
\ne 0$. The nonlocality gives rise to a \textit{modulation instability} and
allows for the existence of several bands of unstable wavevectors 
($\omega_R>0$) \cite{ramazza}.

\begin{figure}[h]
\includegraphics[width=0.23\textwidth]{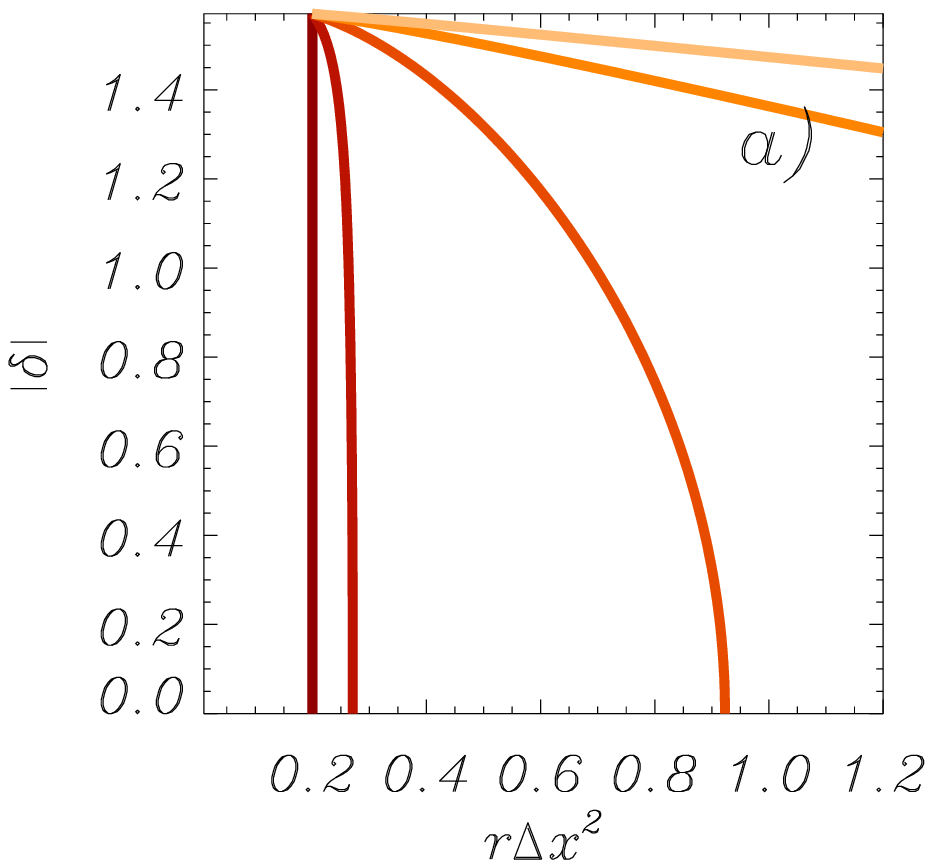}%
\includegraphics[width=0.23\textwidth]{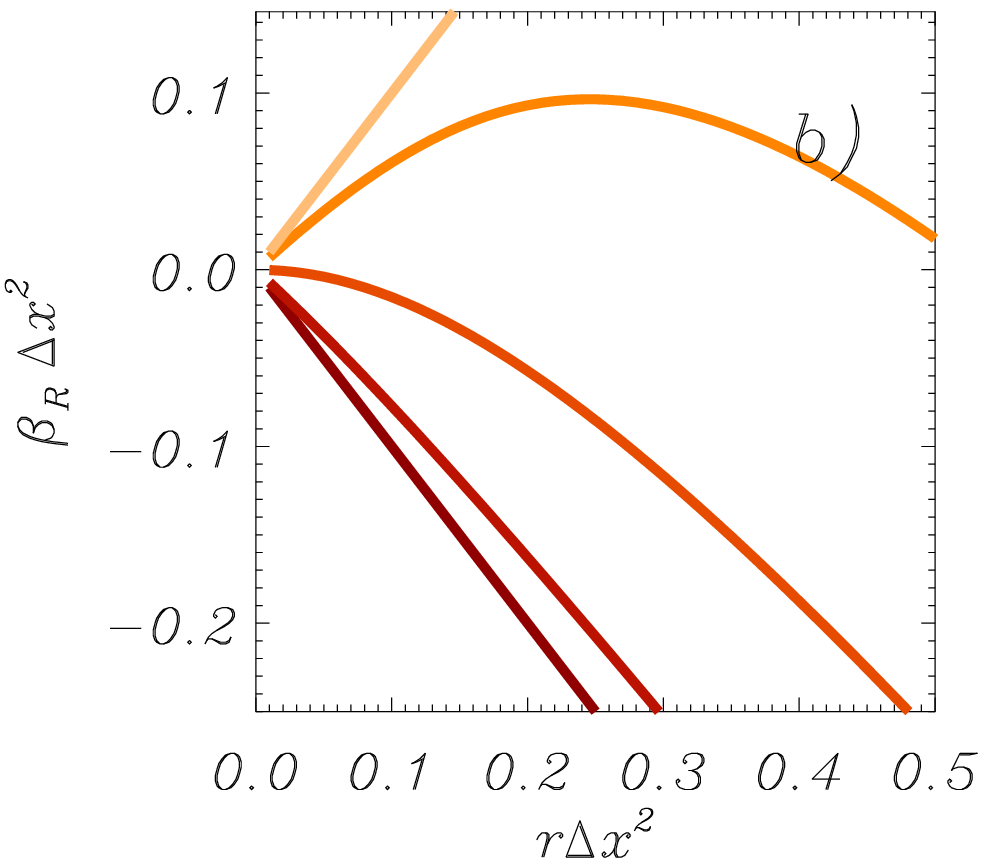} 
\includegraphics[width=0.23\textwidth]{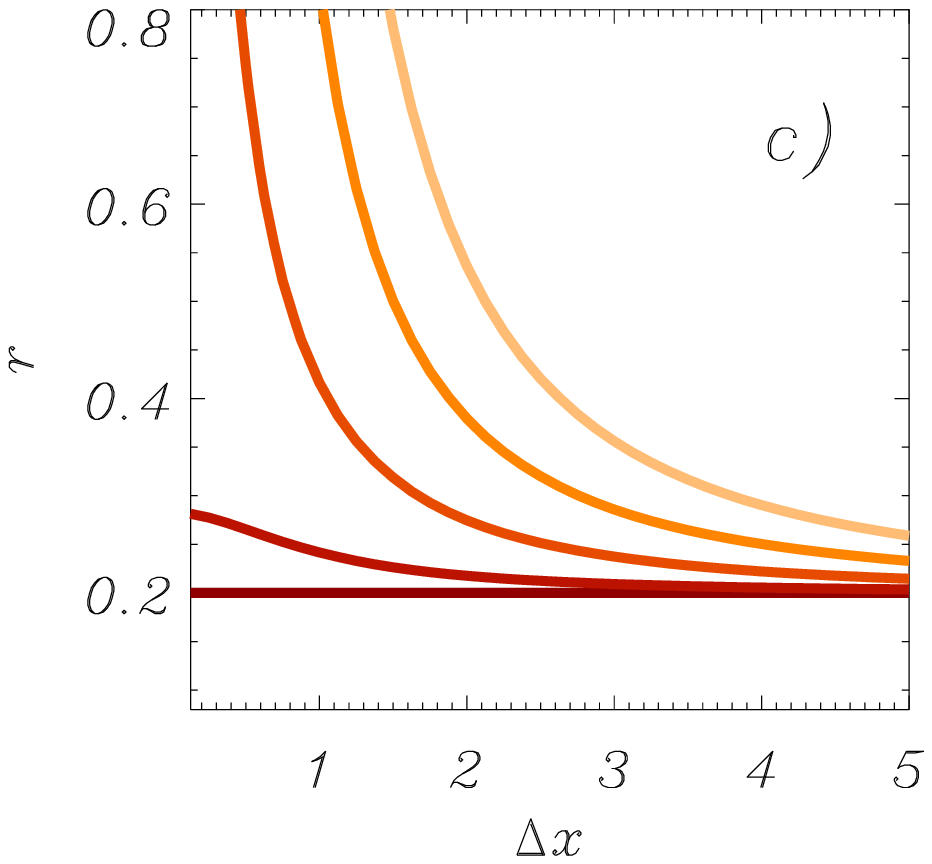}%
\includegraphics[width=0.23\textwidth]{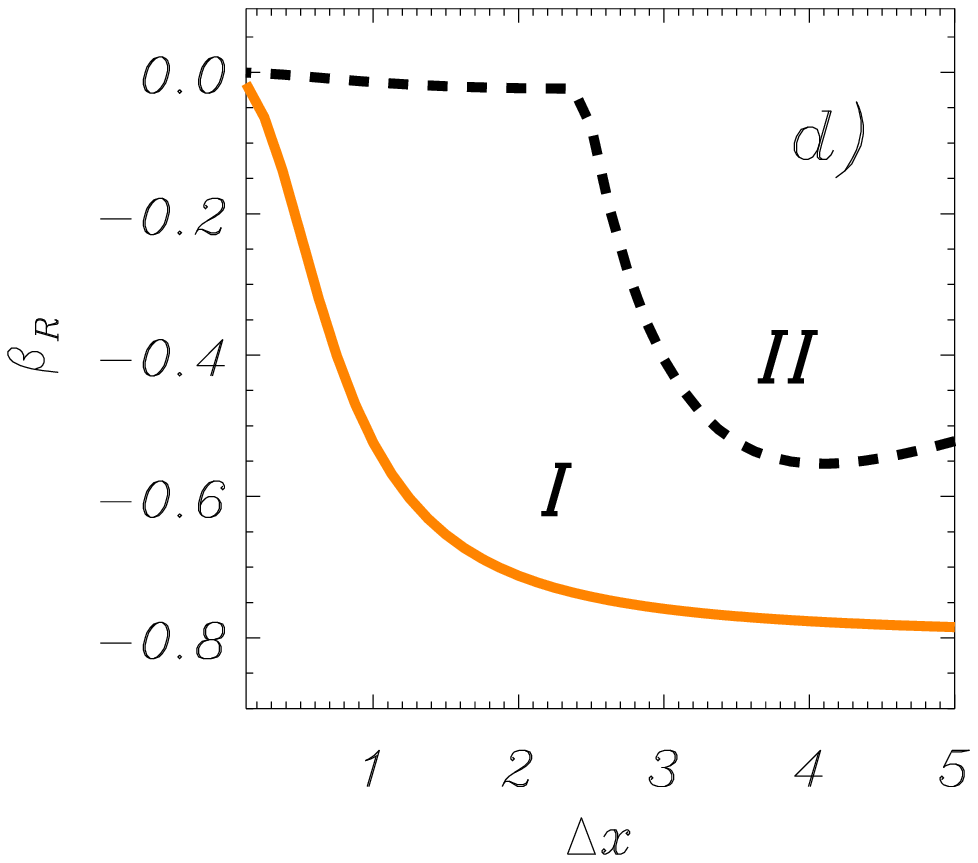} 
\caption{ a) Instability thresholds for $\beta_R\Delta x^2= -0.2$ and for
$\phi=n\pi/4 $ with $n=0,1,2,3,4$ (from dark to light colors). The lowest
threshold is found for  $\phi=0$ and the instability takes place on the right of
the lines. b) Thresholds for $\delta=0.45\pi$ and different feedback $\phi$ as
in (a). c) Thresholds  for $\beta_R= -0.2$ , $\delta=0.45\pi$ and different
feedback $\phi$ as in (a). d) First (continuous line) and second  (dashed line)
thresholds for $\delta=0.45\pi$, $r=0.8$ and $\phi = \pi/2$.}	   
 \label{fig:1}
\end{figure}

The off-axis feedback, besides modulation instability to several bands of
wave\-vectors, provides a wide tunability of the properties of the device and
enables to control the \textit{first threshold}. Inspection of
Eq.~(\ref{ap_disp}) shows that the instability threshold can be expressed as a
function of four relevant parameters, namely $\phi, \delta, r\Delta x^2$, and
$\beta_R\Delta x^2$ (see Figs.~\ref{fig:1}a-b); therefore increasing the shift
size  $\Delta x$ produces on the device the same effect of larger gain
$\beta_R$  and feedback $r$. As a specific effect of the nonlocality, we find
that the relative strength of diffusion and diffraction, $\delta$, also becomes
an effective parameter to control the threshold position.  Indeed, the lowest
gain and feedback thresholds (independently on
 the feedback phase $\phi$) are generally found in the purely diffractive limit
($\delta\sim\pi/2$). 
 The effect of diffusion on the feedback lasing threshold
can be appreciated in Fig.~\ref{fig:1}a: for any not vanishing feedback phase
$\phi$, the threshold value for  the scaled feedback strength $r\Delta x^2$
increases with the diffusion, being independent on the sign of the refractive
index (sign of $\delta$). Both $ \beta_R$ and $r$ can be
increased to cross the laser threshold as shown in Fig.~\ref{fig:1}b, and
--similarly to the case of perfect alignment-- if the feedback is out of phase
then stronger gain is required.   %
For fixed values either of the gain or of the feedback the nonlocality strongly
decreases the threshold values for the gain as well as for the feedback field,
as seen in Figs.~\ref{fig:1}c and d. This can be understood considering that
the  most unstable mode has $k \ne 0$ so that the effect of the nonlocal
coupling is equivalent to a reduction of the feedback dephasing.
Consistently with this interpretation, in the  case of feedback perfectly in
phase with the intracavity field ($\phi=0$) the threshold is independent on the
lateral shift $\Delta x$ because the most unstable mode is the homogeneous one
($k=0$). 

\begin{figure}[h]
\includegraphics[width=0.23\textwidth]{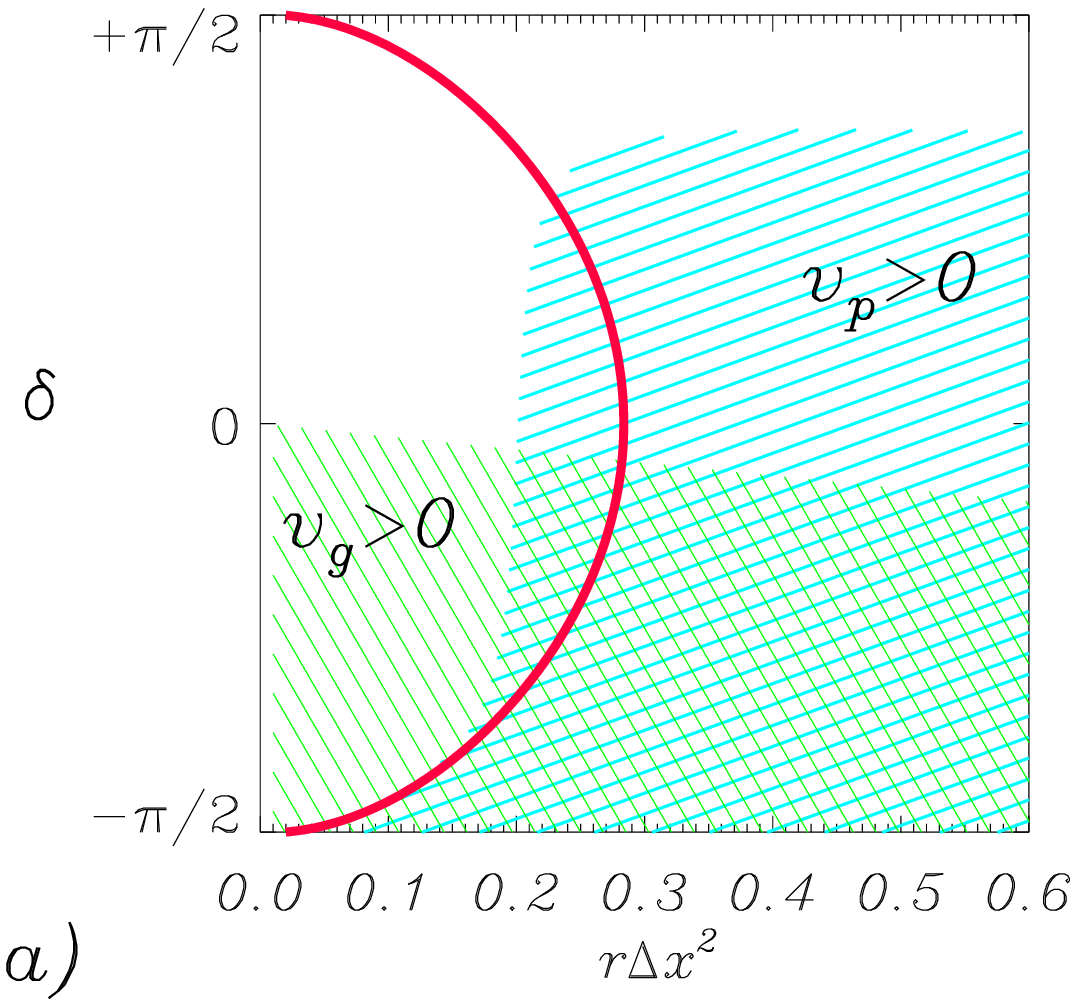}%
\includegraphics[width=0.23\textwidth]{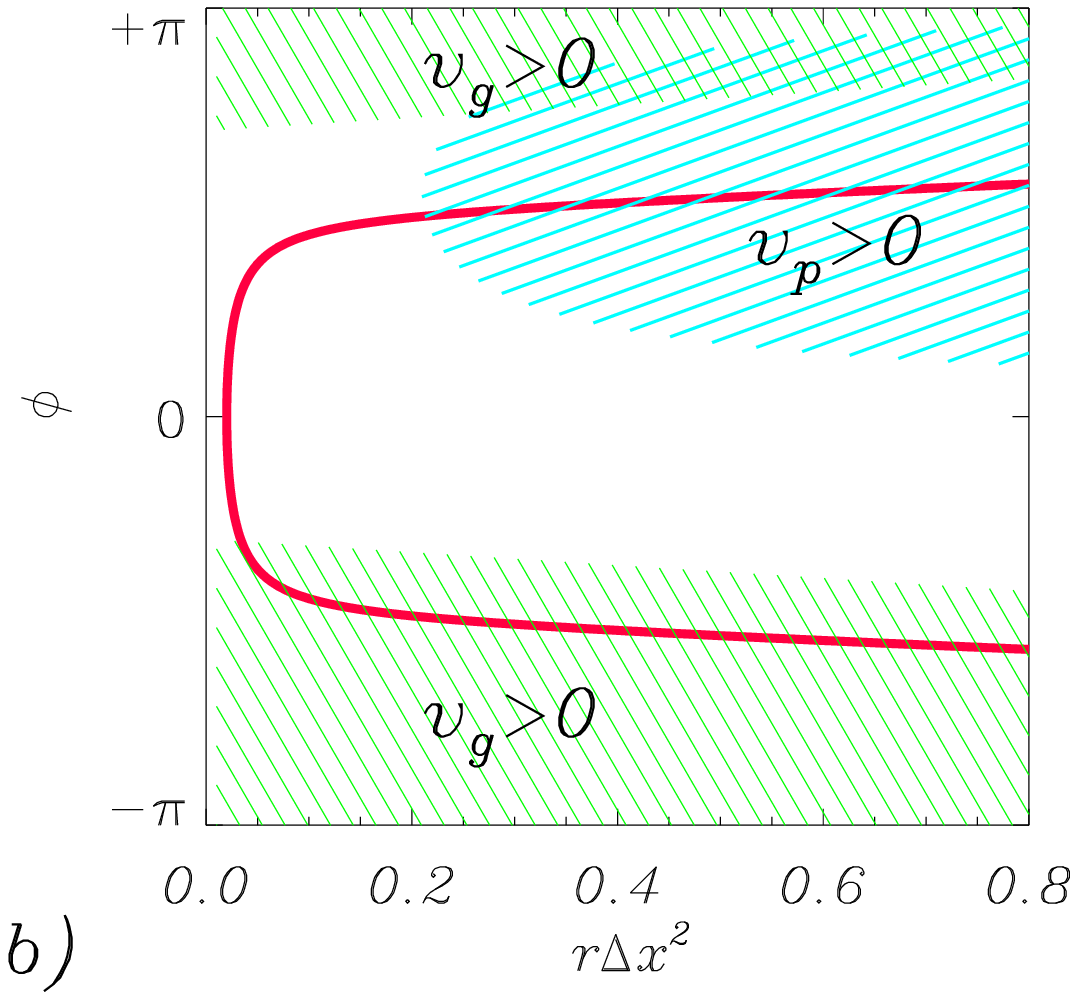} 
 \caption{Sign of  velocities for $\beta\Delta x^2=-0.02-i0.2$, and for $\phi
 =  \pi/2$ (a)  and  $\delta =  0.2\pi$ (b). The dashed regions show where 
 phase and group velocities are positive, while the continuous line marks the
 instability threshold (the system is below threshold on the left sides).
 Negative values of $\delta$ correspond to negative refractive indexes.
 }	   
 \label{fig:2}
\end{figure}

Another effect of the nonlocality concerns
the possibility to  {\sl tune} transverse phase and group velocities 
{\sl independently} from one another. This property enables non mechanical 
steering and spatial chirping of light beams as the high spatial frequencies
can accumulate in the left or right side of the beam. We remark that, 
as for conventional lasers without off-axis
feedback~\cite{Jakobsen},  phase travelling waves are exact solutions of the 
model. Phase and group velocities follow from
Eq.~(\ref{ap_disp}): 
\bnea
v_p &=&-\frac{\omega_I(k)}{k}= k \sin{\delta}  -  
\frac{ \beta_I 
+ r \sin{(k \Delta x + \phi)}}{k} \label{v_p} \\
v_g &=&-\p_{k}\omega_I = 2 {k}  \sin{\delta}  - r \Delta x
\cos{({k}  \Delta x + \phi)}. \label{v_g} 
\enea 
They can be tuned independently because the parameter $\beta_I$ enters only in
the expression of the phase velocity. Evaluation of the velocities for the
critical wavevectors $k_c$ allows us to identify the manifolds in the control
parameter  space that separate regions in which the group and the phase velocity
have the same sign from region in which these velocities have opposite sign. In
particular, the group velocity is null for  $r \Delta x^2=- 2\delta \pm (4n+1)
\pi \mp 2\phi. $ As shown in  Fig.~\ref{fig:2} equal or opposite signs of the
velocities can be observed also $at$ the instability threshold  of the device
(continuous line) depending on the values of $\delta$, $r\Delta x^2$ and $\phi$.
The latter is a promising candidate to tune non mechanically the velocities, for
instance by changing the refractive index in the feedback  loop.

Whenever the group velocity is non null, one has to determine whether amplified
perturbations of the unstable reference state $E_0$ will drift away  (convective
instability), or will  fill the entire system  (absolute instability). The convective
regime is the one where the control of localized light signals is possible. The nature
of the instability  is determined by finding the limit of the Green function of the
linearised system of equations for large time.  The asymptotic local behaviour of the
perturbation is found by generalising the saddle point technique developed in
\cite{papoff05a,zambrini06a} --the details  will be reported elsewhere. In
Fig.~\ref{fig:1}d  we show an example of thresholds of convective (I) and absolute (II)
instabilities; for any choice of parameters there are windows of convective instability
before reaching the lasing thresholds. By using the information in 
Figs.~\ref{fig:1}-\ref{fig:2} and Eqs. (\ref{v_p}-\ref{v_g}) we can determine linear
amplification, direction of propagation and spatial chirping of any light spot in the
transverse plane.

\begin{figure}[h]
\includegraphics[width=0.4\textwidth,clip=true]{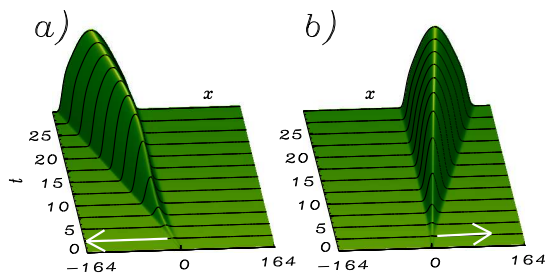}
	\caption{Spatio-temporal diagram for the field intensity $|E|^2$
starting from a small Gaussian perturbation of the vanishing state $E_0$,
 obtained by numerical simulation of Eqs. (\ref{eq:laser}). Parameters: $\mu=0.98$,
$\theta=0.2$, $\delta=0.49\pi$, $r=0.5$, $\Delta x=1$ (coupling each point with a shifted
one on the right) and $\phi=\pi/2$ (a), $\phi=-\pi/2$ (b).}      
	\label{fig:run4}
\end{figure}

In order to check to what extent the linear analysis we reported predicts the
dynamics of the full nonlinear device we consider the standard model for class A
lasers, obtained from Eqs. (\ref{eqs_classB}) with
\bnea \label{eq:laser}
g_1=
-(1+i \theta - N)E ,~~ N=\mu/(1+|E|^2)], 
\enea 
with the usual  parameters  $\theta$ for the detuning with respect to the medium
resonance, and $\mu$ for the pump \cite{laserb}.  The dispersion relation for
the field perturbations around the homogeneous steady state $E_0 = 0$ are given
by Eq.~(\ref{ap_disp}) with $\beta=\mu - 1 - i \theta$.  Numerical simulations 
confirm  the predicted thresholds, in agreement with the
stability diagrams in Fig.~\ref{fig:1}. Moreover, the wavenumbers dynamically
selected  and the velocities are well approximated by those obtained from 
linear dispersion. 
In view of applications it is interesting to see the dynamics of local
perturbation of the homogeneous state: In Fig. \ref{fig:run4} we demonstrate
the ability of steering and amplifying beams in the 
convective region; 
furthermore, one or both the 
signs of phase and group velocities can be changed with the proper parameters choice
(Figs. \ref{fig:run4}a-b), 
consistently with predictions presented in Fig.~\ref{fig:2}. 
Numerical simulations also confirm the possibility of
chirping; the phase of the field shows indeed a spatial dependent modulation.

\begin{figure}[h]
 \includegraphics[width=0.4\textwidth,clip=true]{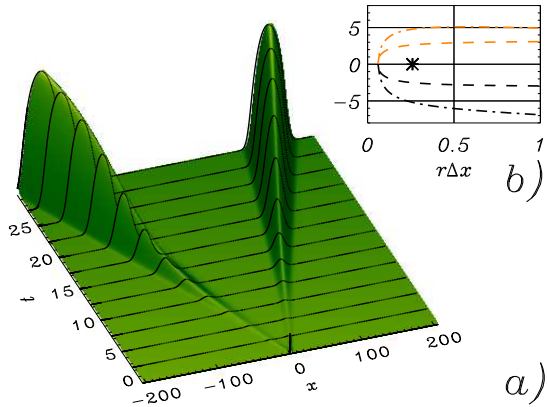}
   \caption{a) Evolution of a Gaussian perturbation as in
   Fig.~\ref{fig:run4} but for $\phi=\pi$. b) Phase
   (dashed lines) and group (dahsed-dotted lines) velocities.  The upper (lower)
   curves are the velocities for  $k_c$ ($-k_c$). For $r\Delta x>0.26$ (star
   point) the homogeneous state is unstable.}	  
 \label{fig:5bis}
\end{figure}

Special attention needs the case $|\phi|=\pi$ where a small spot of light is
amplified and splits in two separate spots travelling in opposite directions as
shown in Fig.~\ref{fig:5bis}a. Both positive and negative wavevectors with
values around the critical ones are selected and then separate moving in
opposite regions of the beam area.  Our analysis for  $|\phi|=\pi$ gives
$\omega_R(k)=\omega_R(-k)$ but, in general, $\omega_I(k)\neq\pm\omega_I(-k)$.
This is important because  in order to see a propagating stripe, for instance
$E\propto \cos(kx+\omega t)$,  it would be necessary to have an antisymmetric 
dispersion $\omega_I(k)$ and the simultaneous instability of  both positive and
negative wavenumbers. This would guarantee that  the interfering waves $k$ and
$-k$ have the $same$   velocities. 
 As shown in Fig.~\ref{fig:5bis}b this is not the case for off-axis feedback:
the phase and group velocities of opposite waves with critical
wavenumbers have $opposite$ signs, and in the diffraction limit
$\delta\rightarrow \pi/2$  both velocities are odd functions
of $k_c$. 
Therefore, even if for  $\phi=\pi$  both $+k_c$ and $-k_c$ are unstable, from
the linear analysis we do not expect intensity stripe patterns above threshold. 
The existence of exact travelling phase patterns  as well as the lack of 
intensity waves are also known in lasers without feedback \cite{Jakobsen}. The
novelty here is the {\it prediction of a state in which  two waves with
wave-vectors $\pm k$ travel apart with opposite velocities}. In spite of the
definite direction associated to the break of reflection symmetry due to two-point  
nonlocality, $both$ transverse direction of propagation are equally 
linearly amplified. 
As shown in Fig.~\ref{fig:5bis}a,
numerical simulations of the model (\ref{eq:laser}) for
$|\phi|=\pi$ fairly agree with these predictions. Even if the linear
amplification of both waves has the same strength, one wave is  nonlinearly
favoured over the other so that a slightly  larger intensity and size of the
packet are found on one side  with respect to the other, depending on the sign
of the shift. As a matter of fact, one mode in the far field is more intense of
the other, similarly to what is found in systems with drift \cite{zambrini-taki}. 
We also note that in this case only the Green function correctly characterises the
convective or absolute nature of the instability. The standard evaluation of the
instability solely in terms of the velocities of the external fronts  of a
perturbation would erroneously describe the convective instability as absolute. We
have seen in fact that here a Gaussian perturbation splits into two wave-packets
with the external fronts moving in opposite directions, as is usually the case for
absolute instabilities, even if the signal eventually decays between the external
fronts.

In conclusion, we have reported a general analysis of the effects of off-axis
feedback in a large class of optical cavities and gain media,  and shown the
threshold dependence on two-point nonlocality, diffusion and  positive as well as
negative diffraction. The possibility to observe  travelling waves at the onset of
the instability in media with fast relaxation  of the polarization is an important
effect of nonlocality, that induces the modulations character of the instability.
We have determined the convective and absolute threshold extending our analysis of
purely diffusive systems \cite{papoff05a}. In presence of nonlocality  phase and
group velocities of optical fields can be easily tuned to parallel or opposite
directions, which enable steering and spatial chirping. Surprisingly, for a
particular  phase of the feedback loop ($\phi=\pi$) we have found the simultaneous
presence of waves  travelling apart. The effect is almost  symmetrical in the
positive and negative directions, even if the  off-axis feedback introduces a
directional coupling in the transverse  plane. The possibility to amplify an
initial spot of light, control its velocity and spatial chirping and even split it
in two counter-propagating signals makes cavities with off-axis feedback a 
promising candidate in view  of  applications in all-optical communications based
on the control of light signals, such as optical triggering, switching, routing,
delay lines, beam recovery and steering and in manipulation of microparticles.
Finally, our theoretical results formally apply to a broad class of devices and
similar effects can be observed for localized perturbations of any nonlocal and
spatially extended system near the onset of oscillations.

Funding from MEC (``Ramon y Cajal" contract and SICOFIB project), from Govern
Balear ("Quantum light in microdevices" project) and from the european project
FunFACS are acknowledged.


\begin{thebibliography}{99}
\bibitem{footnote2}In the case of nonlinear feedback
$[f(|E|^2)E]_{\Delta x}$, the  Eq.~(\ref{ap_disp}) needs to be
considered with  with  $r e^{i \phi}=f(0)$.
\end{thebibliography}
\end{document}